\documentclass[twocolumn,10pt]{asme2e}

\usepackage{mathtools}
\usepackage{array}
\usepackage{epsfig} % for postscript graphics files
\usepackage{dblfloatfix}    % To enable figures at the bottom of page
\usepackage{graphicx}
\usepackage{amsmath}
\usepackage{array}
\usepackage{amssymb}
\usepackage{verbatim}
\usepackage{cleveref}
\usepackage{color}
\usepackage{relsize}
\usepackage{fancyhdr}
\usepackage{graphics}
\usepackage{setspace}
\usepackage{epstopdf}
\usepackage{multirow}
\usepackage{lipsum}
\usepackage{textcomp}
\usepackage{mathtools}
\usepackage{cite}
\usepackage{setspace}
\usepackage{booktabs}
\usepackage{acronym}
\conffullname{ASME 2019 Dynamic Systems and Control Conference (DSCC)}

\confdate{October 8-11}
\confyear{2019}
\confcity{Park City, Utah}
\confcountry{USA}

\papernum{DSCC2019-8969}

\title{Combined Energy and Comfort Optimization of Air Conditioning System in Connected and Automated Vehicles}
\author{Hao Wang\thanks{Address all correspondence to this author.}\\
	{\tensfb Mohammad Reza Amini}\\
	{\tensfb Ziyou Song}\\
	{\tensfb Jing Sun} 
    \affiliation{
	Department of Naval Architecture \& Marine Engineering,\\
	The University of Michigan,\\
	Ann Arbor, MI,\\
    Emails: \{autowang, mamini, ziyou, jingsun\}@umich.edu.
    }	
}

\author{Ilya Kolmanovsky     
    \affiliation{Department of Aerospace Engineering,\\
	The University of Michigan,\\
	Ann Arbor, MI,\\
	Email: ilya@umich.edu.
    }
}

\begin{document}

\maketitle    
%\doublespacing
%%%%%%%%%%%%%%%%%%%%%%%%%%%%%%%%%%%%%%%%%%%%%%%%%%%%%%%%%%%%%%%%%%%%%%
\begin{abstract}
{\it In this paper, we propose a combined energy and comfort optimization (CECO) strategy for the air conditioning (A/C) system of the connected and automated vehicles (CAVs). By leveraging the weather and traffic predictions enabled by the emerging CAV technologies, the proposed strategy is able to minimize the A/C system energy consumption while maintaining the occupant thermal comfort (OTC) within the comfort constraints, where the comfort is quantified by a modified predictive mean vote (PMV) model adapted for an automotive application. A general CECO problem is formulated and addressed using model predictive control (MPC) and weather/traffic previews. Depending on the ways of exploiting the preview information and enforcing the OTC constraint, different MPCs are developed based on solving different variations of the general CECO problem. The CECO-based MPCs are then tested in simulation using an automotive A/C system simulation model (CoolSim) as the virtual testbed. The simulation results show that, over SC03 driving cycle, the proposed CECO-based MPCs outperform the baseline cabin temperature tracking controller, reducing the A/C system energy consumption by up to 7.6\%, while achieving better OTC according to the PMV-based metrics. This energy saving in A/C system translates to 3.1\% vehicle fuel economy improvement. The trade-off between energy efficiency and OTC for different control scenarios is also highlighted.}
\end{abstract}

%%%%%%%%%%%%%%%%%%%%%%%%%%%%%%%%%%%%%%%%%%%%%%%%%%%%%%%%%%%%%%%%%%%%%%
%\begin{nomenclature}
%\entry{CECO}{combined energy and comfort optimization, [-]}
%\entry{CECO-E}{combined energy and comfort optimization with energy saving priority, [-]}
%\entry{CECO-C}{combined energy and comfort optimization with occupant thermal comfort priority, [-]}
%\entry{CECO-CSV}{combined energy and comfort optimization with comfort slack variable, [-]}
%\entry{$\alpha$}{There are two arguments for each entry of the nomemclature environment, the symbol and the definition.}
%\end{nomenclature}
%%%%%%%%%%%%%%%%%%%%%%%%%%%%%%%%%%%%%%%%%%%%%%%%%%%%%%%%%%%%%%%%%%%%%%
\section{Introduction}
The advent of connected and automated vehicle (CAV) technology has created tremendous opportunities from control and optimization perspective to improve mobility, safety, and fuel economy/energy efficiency of the vehicles. With CAV, it is expected that the vehicle control system is able to exploit the vehicle to vehicle (V2V) and vehicle to infrastructure (V2I) communications for more efficient vehicle operation. Notably, most of existing CAV-related research focuses on utilizing information from V2V/V2I or V2X (e.g., traffic signal and traffic flow information) to improve the powertrain system efficiency \cite{Guanetti18, Vahidi18} via eco-driving/platooning and route planning strategies. At the same time, the literature focusing on the vehicle thermal management and co-optimization with traction power is very limited.        

For light-duty vehicles, the power used by the air conditioning (A/C) system usually represents the most significant thermal load in summer. It has been shown that the A/C thermal load may lead to dramatic vehicle range reduction, especially for the vehicles with electrified powertrains \cite{Jeffers2015,Rask2014}. Besides its noticeable impact on vehicle range reduction, the performance of A/C system also has a direct influence on occupant comfort and customer satisfaction. Uncomfortable cabin thermal conditions can increase the stress for the vehicle passengers, thereby inducing higher chances of traffic accidents \cite{Daanen2003}. 
%Interactions between vehicle energy consumption and occupant thermal comfort (OTC) render concurrent optimization of the energy and the OTC essential. 
For a conventional A/C system with belt-driven compressor, the A/C load directly affects the operation of the internal combustion engine (ICE). In this case, the coordinated energy management of the A/C and the ICE has been studied \cite{HKhayyam2011,QZhang2016}, and the corresponding thermal comfort implications have been investigated more recently in the context of a multi-objective proportional-integral (PI) controller \cite{XYan2018}. However, the V2X information has not been exploited in these works for the A/C energy management. Moreover, as the vehicle powertrain is becoming more electrified, such mechanical coupling and associated coordination between A/C and ICE will disappear as the A/C compressor becomes electric-driven and with power supplied from the on-board battery. Our previous publications on automotive A/C energy management \cite{Hao18, Reza19, Reza19_2} have exploited the sensitivity of A/C system efficiency to vehicle speed and vehicle speed preview for reducing energy consumption. The average cabin temperature was used as the comfort metric in \cite{Hao18, Reza19, Reza19_2}. However, as shown in \cite{Martinho2003}, the OTC may also depend on other variables. More recently, we have introduced the discharge air cooling power (DACP) in  \cite{Hao19}, which reflects the dependence of the OTC on the vent air temperature, the air velocity, and the average cabin temperature.  Other variables, such as solar radiation and humidity, are however not captured by the DACP metric. 

In order to comprehensively represent the OTC, we utilize the predicted mean vote (PMV) model \cite{Hoof08,ISO7730} that is well-known and has been adopted by the heating ventilation and air conditioning (HVAC) community. This PMV model was originally developed for evaluating the indoor thermal comfort for building HVAC systems using the data from human experiments in a thermal chamber. The PMV model has been previously applied to temperature control in buildings \cite{Croitoru15, Ku15, Chen15}. Comfort-based climate control for automotive applications using PMV has also been of recent interest, see \cite{Farzaneh08, Brusey18, XYan2018, SSchaut19}.

In this paper, we propose a combined energy and comfort optimization (CECO) strategy for the A/C system in vehicles with electrified powertrains. The comfort evaluation is performed based on a modified PMV model. Compared with the existing PMV-based control designs \cite{Farzaneh08, XYan2018}, which directly adopted the PMV model developed for building applications, our modified PMV model accounts for the special characteristics of automotive applications by explicitly accounting for the solar radiation and vent air temperature. Moreover, the proposed CECO strategy leverages the weather and traffic predictions made available via V2X communication for improving the energy efficiency, while maintaining the OTC within the specified constraints. A nonlinear model predictive control (NMPC) approach is then pursued in order to exploit the preview information and handle constraints. Using a high-fidelity automotive A/C system model (CoolSim) \cite{Kiss13}, the effectiveness of the proposed CECO strategy is demonstrated versus a baseline control strategy that tracks a constant cabin air temperature set-point. In addition, the trade-off between the energy efficiency and the OTC is illustrated through different control case studies.

The rest of the paper is organized as follows. The A/C system in an electrified powertrain, and the corresponding control-oriented model are described in Sec.~\ref{sec:1}. Next, the modified PMV model and the OTC constraints are described in Sec.~\ref{sec:2}. In Sec.~\ref{sec:3}, detailed CECO problem formulation is presented. Simulation results are reported in Sec.~\ref{sec:4}. Eventually, conclusions are given in Sec.~\ref{sec:5}.\vspace{-0.5cm}
%%%%%%%%%%%%%%%%%%%%%%%%%%%%%%%%%%%%%%%%%%%%%%%%%%%%%%%%%%%%%%%%%%%%%%
\section{Air Conditioning (A/C) System in an Electrified Powertrain}\label{sec:1}%
\begin{figure}[h!]
	\begin{center}
		\includegraphics[width=9cm]{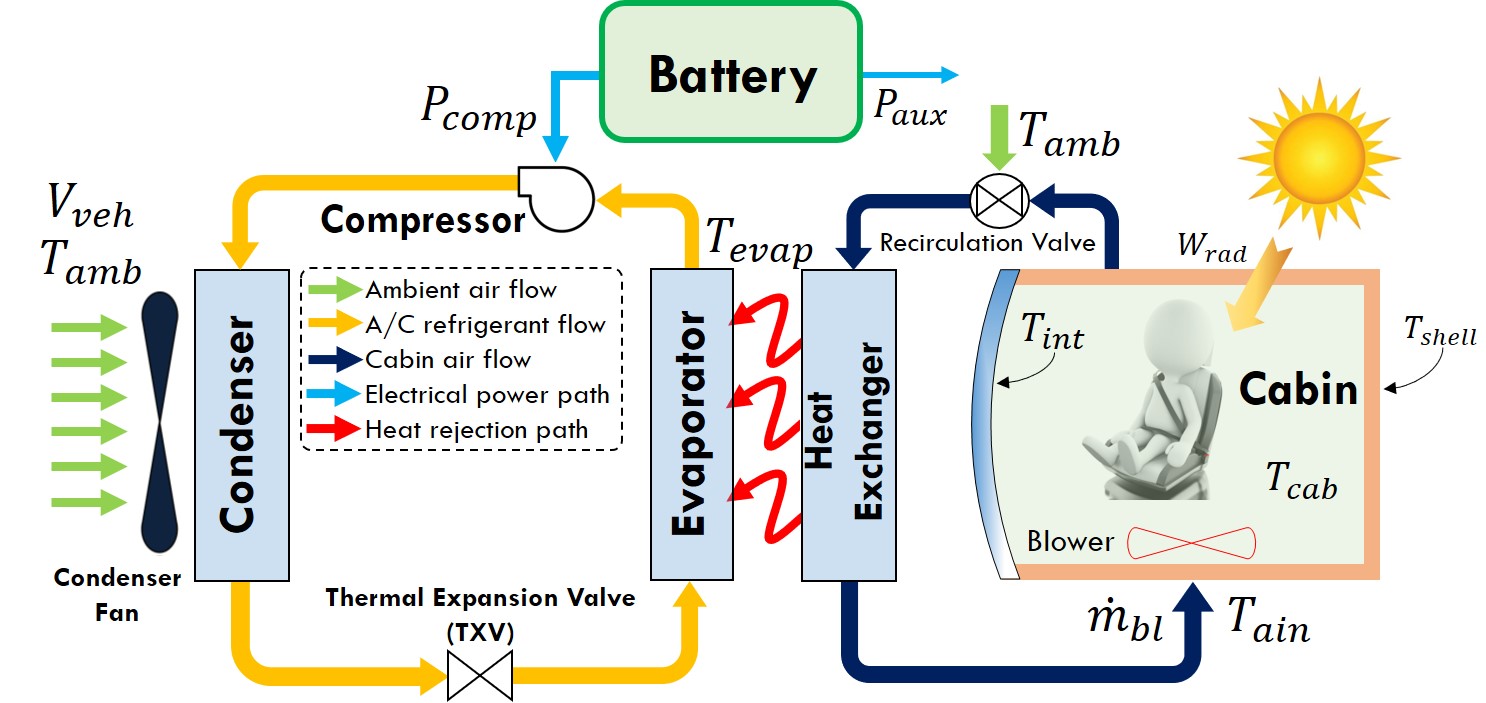} \vspace{-0.5cm}   % The printed column width is 8.4 cm.
	\end{center}\vspace{-0.5cm}
	\caption{Schematic of the A/C system in an electrified powertrain.}
	\label{fig_AC_schematic} 
\end{figure}

Fig.~\ref{fig_AC_schematic} provides a schematic of a typical A/C system in an electrified powertrain in which an onboard battery supplies the power to the major power consumers in the A/C system, namely, the compressor ($P_{comp}$) and the other auxiliaries ($P_{aux}$) including the condenser fan and the blower. There are two major loops of flow in the A/C system, the refrigerant loop (RL) shown in yellow lines and the cabin air loop (CAL) shown in blue lines. In practice, depending on the cooling power demand from the CAL, the actuators in the RL including the compressor, the condenser fan, and the thermal expansion valve, etc., are coordinated to maintain the evaporator wall temperature ($T_{evap}$) within the desired and safe range. From the comfort perspective, there are many variables within the CAL that will influence the OTC such as the average cabin air temperature ($T_{cab}$), the cabin interior (e.g., seats and panels) temperature ($T_{int}$), the vent air temperature ($T_{ain}$), the air flow rate ($\dot{m}_{bl}$), and the solar radiation ($W_{rad}$) \cite{Martinho2003}. These variables will be considered in a more comprehensive OTC model described in the next section. The variables $V_{veh}$, $T_{amb}$, $T_{shell}$ represent the vehicle speed, ambient air temperature, and cabin shell temperature, respectively. \vspace{-0.5cm}
 \vspace{-0.5cm} 

\begin{figure}[h!]
	\begin{center}
		\includegraphics[width=6cm]{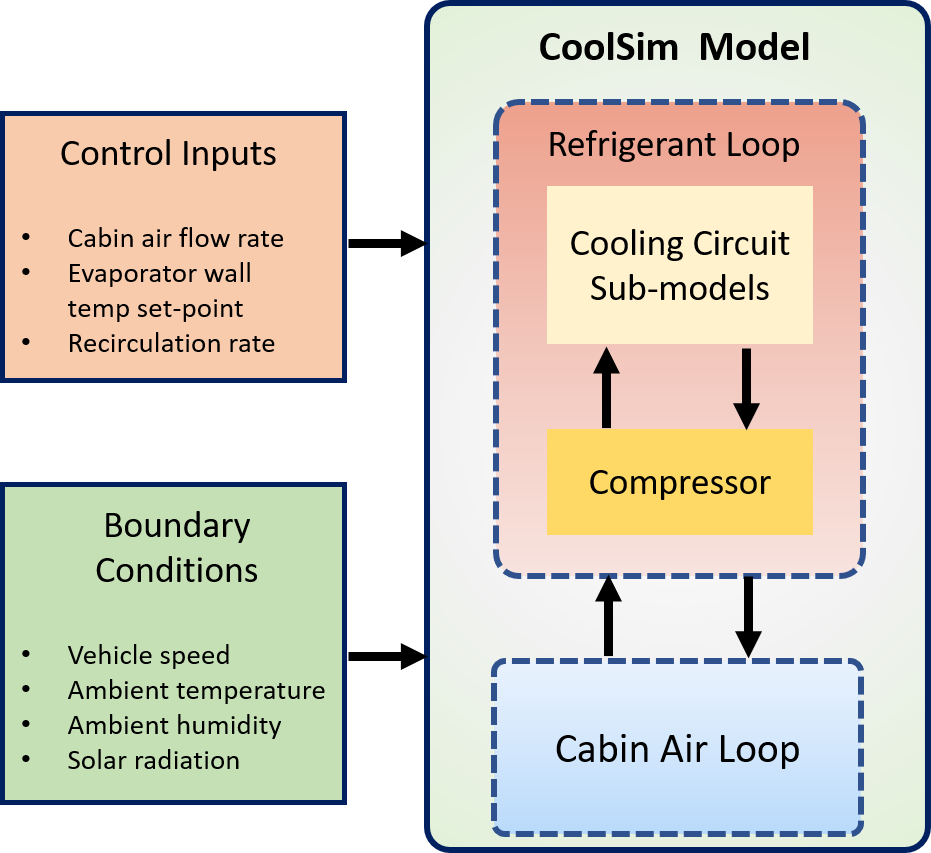} \vspace{-0.3cm}   % The printed column width is 8.4 cm.
	\end{center}
	\caption{Schematic of the CoolSim model.}\vspace{-0.5cm}
	\label{fig_CoolSim_schematic} 
\end{figure}
The detailed physical modeling of this A/C system and especially the modeling of the RL is complicated \cite{QZhang2016_2}. In order to validate the control design, we adopt a high-fidelity CoolSim model developed by the National Renewable Energy Lab (NREL) \cite{Kiss13}. Its architecture is shown in Fig.~\ref{fig_CoolSim_schematic}. Simulations of CoolSim model have revealed \cite{Hao18} that the A/C efficiency increases as vehicle speed increases. This is attributed to the condenser dissipating the heat more efficiently as the ram air speed increases. This sensitivity will also be exploited in the CECO strategy developed in this paper to facilitate the efficient and comfortable A/C operation. A control-oriented discrete-time model of the A/C system has been developed in \cite{Hao18}, and has the following form:\vspace{-0.3cm} 
\begin{eqnarray}
	\label{eqn:ACModel_Tcab}
	T_{cab}(k+1) & = & f_{T_{cab}}(k)=T_{cab}(k)+\gamma_1(T_{int}(k)-T_{cab}(k))\\
	&+&\gamma_2(T_{shell}(k)-T_{cab}(k)) \nonumber\\ &+&\gamma_3(T_{ain}(k)-T_{cab}(k))\dot{m}_{bl}(k)+\tau_{1},\nonumber\\
	\label{eqn:ACModel_Tevap}
	T_{evap}(k+1)&=&f_{T_{evap}}(k)=\gamma_4T_{evap}(k)\\
    &+&\gamma_5(T_{evap}(k)-T_{evap}^{s.p.}(k))+\tau_{2},\nonumber\\
	\label{eqn:ACModel_Tain}
	T_{ain}(k)&=&\gamma_6T_{evap}(k)+\gamma_7\dot{m}_{bl}(k)+\tau_{3},
\end{eqnarray} 
where the temperatures are defined as indicated in Fig.~\ref{fig_AC_schematic} and have units of $K$ \cite{Hao18}. Two control inputs are the air flow rate through the blower, $\dot{m}_{bl}$ (in $kg/s$), and the evaporator wall temperature set-point, $T_{evap}^{s.p.}$ (in $K$), respectively. The constant parameters ($\gamma_1, ..., \gamma_7$ and $\tau_1, ..., \tau_3$) are identified and validated versus the CoolSim outputs. Furthermore, the compressor power ($P_{comp}$) and the auxiliary power ($P_{aux}$) can also be estimated based on the inputs and states of the model (\ref{eqn:ACModel_Tcab})-(\ref{eqn:ACModel_Tain}). \vspace{-0.5cm} 

\section{Occupant Thermal Comfort (OTC) Model}\label{sec:2} 
In this section, we present the OTC model based on the modified PMV which accounts for multiple factors, including solar radiation and vent air temperature.\vspace{-0.25cm}

\subsection{Original PMV model}
In the original PMV model for indoor spaces described in \cite{ISO7730,Croitoru15}, the PMV index computation exploits the following heat balance equation,\vspace{-0.25cm}
\begin{eqnarray}
\label{eqn:HBE_orignal}
M-W_{mech}=H+E_c+C_{res}+E_{res},
\end{eqnarray}
where $M$ and $W_{mech}$ represent the metabolic rate of the occupant and the effective mechanical power due to work performed by the occupant, respectively. The terms on the right of (\ref{eqn:HBE_orignal}), $H$, $E_c$, $C_{res}$, and $E_{res}$ represent the dry heat loss, the evaporative heat exchange at skin, the respiratory convective heat exchange, and the respiratory evaporative heat exchange, respectively. All these variables are in units of $W/m^2$. When Eqn. (\ref{eqn:HBE_orignal}) holds, the best OTC level is achieved. Otherwise, the occupant feels either warm or cold. Inspired by this heat balance equation, the PMV index is calculated by \cite{Hoof08,ISO7730}:\vspace{-0.35cm}
\begin{eqnarray}
\label{eqn:PMV_original}
y_{PMV}^{*}&=&(0.303e^{-0.036M}+0.028)[(M-W_{mech}) \nonumber\\&-&(H+E_c+C_{res}+E_{res})],\\
\text{where},\nonumber\\
\label{eqn:H}
H&=&3.96\cdot 10^{-8}f_{cl}[(T_{cl}+273)^4-(T_{mr}+273)^4]\nonumber \\&+&f_{cl} h_c(T_{cl}-T_a),\\
E_c&=&3.05\cdot 10^{-3} [5733-6.99\cdot (M-W_{mech})-p_a] \nonumber\\&+&0.42 (M-W_{mech}-58.15),\\
C_{res}&=&0.0014M(34-T_a),\\
E_{res}&=&1.7\cdot10^{-5}M(5867-p_a),\\
T_{cl}&=&35.7-0.0275(M-W_{mech})-I_{cl}\{ M-W_{mech}\nonumber\\&-&3.05[5.73-0.007(M-W_{mech})-p_a]\nonumber \\&-&0.42(M-W_{mech}-58.15)-0.0173M(5.87-p_a)\nonumber \\ &-&0.0014M(34-T_a)\},\\
h_c&=&\left\{ \begin{array}{ll}
2.38|T_{cl}-T_{a}|^{0.25}, & \mbox{$2.38|T_{cl}-T_{a}|^{0.25}>12.1\sqrt{V_{air}}$}, \hspace{+0.5cm}\\
12.1\sqrt{V_{air}}, & \mbox{$2.38|T_{cl}-T_{a}|^{0.25}\leq12.1\sqrt{V_{air}},$}\end{array}\right. \nonumber\\
\label{eqn:fcl}
f_{cl}&=&\left\{ \begin{array}{ll}
1.05+0.645I_{cl}, & \mbox{$I_{cl}>0.078$}, \hspace{+0.5cm}\\
1.00+1.29I_{cl}, & \mbox{$I_{cl}\leq 0.078,$}\end{array}\right.
\end{eqnarray}
with $T_{cl}$, $T_{mr}$, $T_a$ being the cloth surface temperature, the mean radiant temperature, and the air temperature (in $^oC$), respectively. The constants (or variables) $f_{cl}$, $h_c$, $I_{cl}$, $p_a$, and $V_{air}$  represent the clothing surface area factor, the convective heat transfer coefficient (in $W/(m^2\cdot K)$), the clothing insulation (in $m^2\cdot K/W$), the partial water vapor pressure (in $Pa$), and the relative air velocity (in $m/s$), respectively. Note that this model represents the comfort of an average people in population level. The model coefficients of this empirical model may change when applied to a specific individual.   \vspace{-0.5cm}

\subsection{Modified PMV model for automotive applications}
In automotive applications, the passengers are subject to direct solar radiation and their thermal sensations are also influenced by the vent air velocity and temperature since they sit close to the vents. To account for these effects, modifications to the original PMV model are now proposed. Firstly, the heat balance equation (\ref{eqn:HBE_orignal}) is modified as\vspace{-0.3cm} 
\begin{eqnarray}
\label{eqn:HBE}
M+W_{rad}=H+E_c+C_{res}+E_{res},
\end{eqnarray}
where $W_{rad}$ represents the effective solar radiation power in the unit of $W/m^2$ and we assume $W_{mech}=0$ since there is no mechanical work associated with the occupant sitting inside the cabin. Secondly, the PMV index computation (\ref{eqn:PMV_original}) is modified to the following form:\vspace{-0.3cm} 
\begin{eqnarray}
\label{eqn:PMV}
y_{PMV}&=&(0.303e^{-0.036M}+0.028)[(M+W_{rad}) \nonumber\\&-&(H+E_c+C_{res}+E_{res})],
\end{eqnarray}
where  $H$, $E_c$, $C_{res}$, and $E_{res}$ are evaluated based on (\ref{eqn:H})-(\ref{eqn:fcl}) and with \vspace{-0.5cm}
\begin{eqnarray}
\label{eqn:Ta}
T_{a}&=&\alpha_1T_{cab}+\alpha_2T_{ain},
\end{eqnarray}
where $\alpha_1$ and $\alpha_2$ are the parameters introduced to account for the impact of $T_{ain}$. As compared to the original PMV model used in \cite{Farzaneh08, XYan2018}, we introduced the new input $W_{rad}$ to account for the solar radiation impact on the OTC and we combined the impacts of $T_{cab}$ and $T_{ain}$ in Eqn.~(\ref{eqn:Ta}) instead of using $T_a=T_{cab}$. In this work, several assumptions have been made in the $y_{PMV}$ evaluation:
\begin{enumerate}
	\item $W_{rad}$ is time-varying depending on average solar radiation, cloud coverage, vehicle orientation, etc. It is assumed to be known via V2X communications, e.g., using the approach in \cite{Ilya10}. In the simulations, $W_{rad}$ trajectory over the driving cycle is specified to qualitatively demonstrate the solar radiation impact on the OTC (i.e., the occupant tends to feel hotter as $W_{rad}$ increases); 
	\item Similar to \cite{Brusey18}, we use the cabin interior temperature to represent the mean radiant temperature (i.e., $T_{mr}=T_{int}$), which is mainly used for capturing the radiative heat transfer of the human body to the cabin;
	\item $V_{air}$ is assumed to be only affected by $\dot{m}_{bl}$ so that $V_{air}$ may be directly controlled. Furthermore, there is a prescribed linear mapping between $\dot{m}_{bl}$ and $V_{air}$;
	\item For simplicity, humidity control is not considered in current MPC design, therefore $p_a$ is assumed to be constant ($1700$ $Pa$) for the $y_{PMV}$ evaluation.
\end{enumerate}

Note that according to the original definition of the PMV index, $y_{PMV}=0$, where $y_{PMV}$ is defined by Eqn.~(\ref{eqn:PMV}), represents the best comfort level as the heat balance is achieved in Eqn.~(\ref{eqn:HBE}). The occupant feels warm or cold depending on whether $y_{PMV}$ is positive or negative, respectively. In Table~\ref{Tbl:OTS}, the PMV-based thermal sensation level is determined according to \cite{ISO7730}. \vspace{-0.5cm}
\begin{table}[h!]
	\caption{PMV-based occupant thermal sensation level.}
	\begin{center}
	\label{Tbl:OTS}
	\begin{tabular}{|c|c|}
		\hline
		$y_{PMV}$ & Thermal sensation level \\ \hline
		3         & Hot                     \\ \hline
		2         & Warm                    \\ \hline
		1         & Slightly warm           \\ \hline
		0         & Neutral                 \\ \hline
		-1        & Slightly cool           \\ \hline
		-2        & Cool                    \\ \hline
		-3        & Cold                    \\ \hline
	\end{tabular}\vspace{-0.65cm}
\end{center}
\end{table}

\subsection{Occupant thermal comfort (OTC) constraints}
\begin{figure}[h!]
	\begin{center}
	\includegraphics[width=7.0cm]{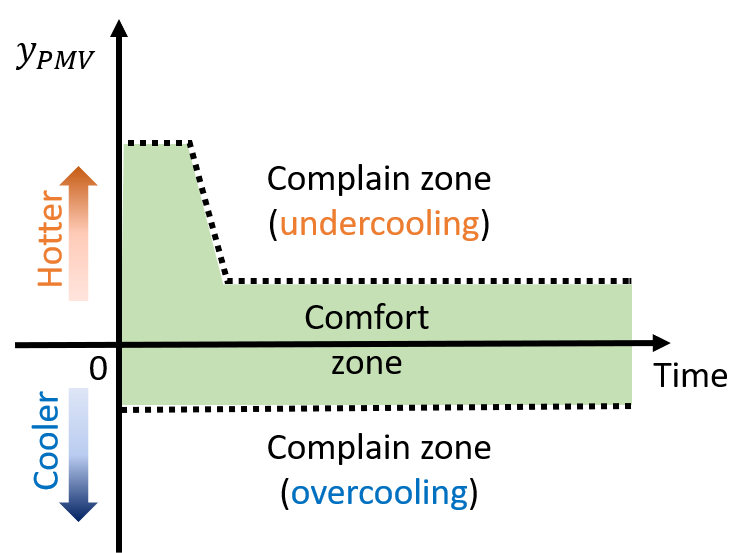} \vspace{-0.3cm}   % The printed column width is 8.4 cm.
	\end{center}
	\caption{Illustration of the OTC constraints.}
	\label{fig_OTC_criteria} 
\end{figure}

As illustrated in Fig.~\ref{fig_OTC_criteria}, the upper and lower bounds on $y_{PMV}$ shown in the dotted black lines are assumed to be prescribed, which may depend on occupant's personal cooling preference. The comfort zone is defined as the region between the upper and lower bounds. The region complimentary to the comfort zone is referred to as the complain zone. We further define the case when $y_{PMV}$ is above the upper bound as undercooling and the case when $y_{PMV}$ is below the lower bound as overcooling. In the cooling case studied in this paper, the upper bound on $y_{PMV}$ is time-varying to accommodate the realistic response of the A/C system in summer as it usually takes some time to cool down the cabin to the comfort level. \vspace{-0.85cm}
%This initial cool-down period is acceptable only within the limited time. The OTC criteria is defined to (i) stay within the comfort zone and (ii) operate around $y_{PMV}=0$, where the former one is hard-constrained. This criteria will be used to formulate the CECO problem and  design the corresponding MPC controllers. 
\vspace{-1cm}

\section{Combined Energy and Comfort Optimization (CECO) Problem Formulation}\label{sec:3}
\subsection{General CECO problem formulation}
\label{sec:3_1}
\begin{figure}[h!]
	\begin{center}
		\includegraphics[width=7.0cm]{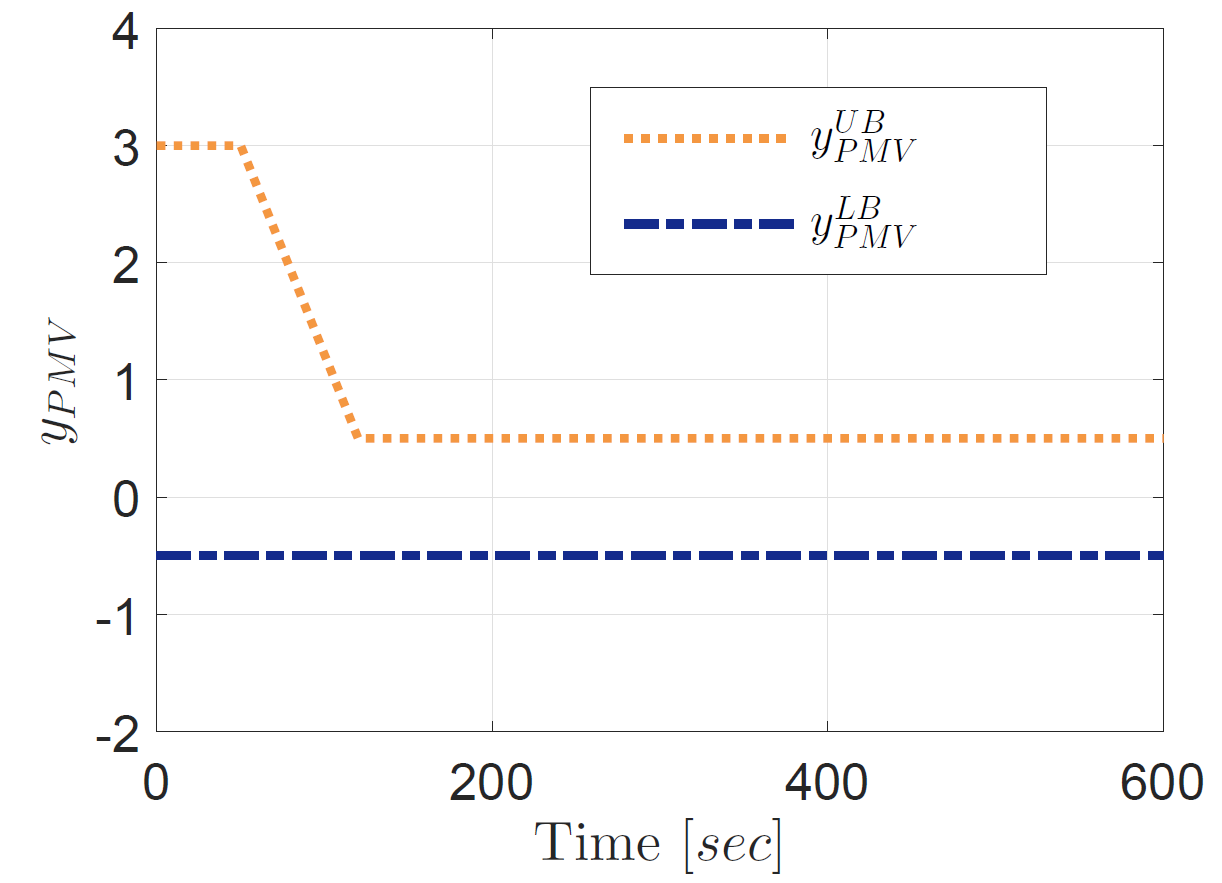} \vspace{-0.3cm}   % The printed column width is 8.4 cm.
	\end{center}\vspace{-0.3cm}  
	\caption{$y_{PMV}^{UB}$ and $y_{PMV}^{LB}$ over SC03 driving cycle.}
	\label{fig_pmv_bound} 
\end{figure}

The general CECO problem is formulated as follows with the objectives of minimizing the energy consumption while maintaining $y_{PMV}$ within comfort zone:\vspace{-0.3cm}
\begin{equation} 
\begin{aligned}\label{eqn:CECO}
& \min_{\substack{\dot{m}_{bl}(\cdot|k)\\T_{evap}^{s.p.}(\cdot|k)}} && \sum_{i=0}^{N_p}  \Big\{\begin{gathered} P_{comp}(i|k)+P_{bl}(i|k) \end{gathered} \Big\},\\
& \text{s.t.}
& & T_{cab}(i+1|k)=f_{T_{cab}}(i|k),~{i=0,\cdots,N_p},\\
&
& & T_{evap}(i+1|k)=f_{T_{evap}}(i|k),~{i=0,\cdots,N_p},\\
&
& &y_{PMV}^{LB}(i|k)\leq y_{PMV}(i|k)\leq y_{PMV}^{UB}(i|k),~{i=0,\cdots,N_p},\\
&
& &T_{evap}^{LB}(i|k)\leq T_{evap}(i|k)\leq T_{evap}^{UB}(i|k),~{i=0,\cdots,N_p},\\
&
& &0.05~kg/s \leq \dot{m}_{bl}(i|k)\leq 0.17~kg/s,~{i=0,\cdots,N_p-1},\\
& 
& &3^oC\leq T_{evap}^{s.p.}(i|k)\leq 10^oC,~{i=0,\cdots,N_p-1},\\
&
& & T_{cab}(0|k)=T_{cab}(k),~T_{evap}(0|k)=T_{evap}(k),%\\
\end{aligned}
\end{equation}
where $(i|k)$ denotes the predicted value of the corresponding variable at time instant $k+i$ when the prediction is made at the time instant $k$, $N_p$ represents the prediction horizon, the overall energy consumption of the A/C system is determined by the sum of compressor ($P_{comp}$) and blower ($P_{bl}$) powers in the cost function, $f_{T_{cab}}(i|k)$ and $f_{T_{evap}}(i|k)$ represent the major system dynamics as defined in Eqns.~(\ref{eqn:ACModel_Tcab}) and (\ref{eqn:ACModel_Tevap}), $y_{PMV}^{LB}$ and $y_{PMV}^{UB}$ are the lower and upper bounds on $y_{PMV}$, and $T_{evap}^{LB}$ and $T_{evap}^{UB}$ are the lower and upper bound on $T_{evap}$, which account for the system operating limits. The lower and upper bounds on $y_{PMV}$ applied in the simulation case studies over SC03 driving cycle are illustrated in Fig.~\ref{fig_pmv_bound}. The upper and lower bounds on $\dot{m}_{bl}$ and $T_{evap}^{s.p.}$ are determined based on our particular A/C system operating limits. Note that  in PMV-related studies \cite{Freire08}, the comfort zone typically corresponds to $y_{PMV}^{LB}=-0.5$ and $y_{PMV}^{UB}=0.5$.\vspace{-0.75cm}
\vspace{-1.2cm}

\subsection{CECO with energy priority (CECO-E) and with comfort priority (CECO-C)}
Based on the general CECO problem formulation in Sec.~\ref{sec:3_1}, different variations in the controller design can be considered according to different objectives. To accommodate the trade-off between the energy consumption and the OTC, a modified cost function as compared to the one in (\ref{eqn:CECO}) is proposed as follows,\vspace{-0.5cm}  
\begin{equation} 
\begin{aligned}\label{eqn:CECO_EC}
	& \min_{\substack{\dot{m}_{bl}(\cdot|k)\\T_{evap}^{s.p.}(\cdot|k)}} && \sum_{i=0}^{N_p}  \Big\{\begin{gathered} P_{comp}(i|k)+P_{bl}(i|k) \end{gathered}+\gamma\cdot y_{PMV}^2(i|k) \Big\},
\end{aligned}
\end{equation}
where $\gamma$ is the penalty on the deviation from $y_{PMV}=0$, which represents the ideal OTC level. We further refer to the case when $\gamma=0$ as \textbf{CECO with energy priority (CECO-E)} and refer to the case when $\gamma$ equals to a large positive number (i.e., $10^5$ in our case) as \textbf{CECO with comfort priority (CECO-C)}. Note that CECO-C is expected to consume more energy for providing better OTC level compared with CECO-E. 

\subsection{CECO with intelligent online constraint handling (CECO-IOCH)}\vspace{-0.55cm}  
As discussed in the previous section, better OTC can be achieved by adding a penalty term to the cost function of the general CECO problem, however, this may unnecessarily increase energy consumption. Here, we propose a more energy efficient approach for improving the OTC, which specifically leverages the vehicle speed preview. This approach is implemented by solving the following variation of the CECO problem with intelligent online constraint handling (IOCH), which is designated as \textbf{CECO-IOCH}. \vspace{-0.3cm}  
\begin{equation} 
\begin{aligned}\label{eqn:CECO-IOCH}
& \min_{\substack{\dot{m}_{bl}(\cdot|k)\\T_{evap}^{s.p.}(\cdot|k)\\\epsilon(\cdot|k)}} && \sum_{i=0}^{N_p} \Bigg\{ \begin{gathered} P_{comp}(i|k)+P_{bl}(i|k)+\beta\Big(\frac{\eta_{AC}(i|k)-1}{\epsilon(i|k)+\xi}\Big) \end{gathered}\Bigg\},\\
& \text{s.t.}
& & T_{cab}(i+1|k)=f_{T_{cab}}(i|k),~{i=0,\cdots,N_p},\\
&
& & T_{evap}(i+1|k)=f_{T_{evap}}(i|k),~{i=0,\cdots,N_p},\\
&
& &y_{PMV}^{LB}(i|k)\leq y_{PMV}(i|k)\leq y_{PMV}^{UB}(i|k)-\epsilon(i|k),~{i=0,\cdots,N_p},\\
&
& &T_{evap}^{LB}(i|k)\leq T_{evap}(i|k)\leq T_{evap}^{UB}(i|k),~{i=0,\cdots,N_p},\\
&
& &0.05~kg/s \leq \dot{m}_{bl}(i|k)\leq 0.17~kg/s,~{i=0,\cdots,N_p-1},\\
& 
& &3^oC\leq T_{evap}^{s.p.}(i|k)\leq 10^oC,~{i=0,\cdots,N_p-1},\\
& 
& &0 \leq \epsilon(i|k)\leq \epsilon^{UB},~{i=0,\cdots,N_p-1},\\
&
& & T_{cab}(0|k)=T_{cab}(k),~T_{evap}(0|k)=T_{evap}(k),%\\
\end{aligned}
\end{equation}

\begin{figure*}[!h]
	\begin{center}
		\includegraphics[scale=0.5]{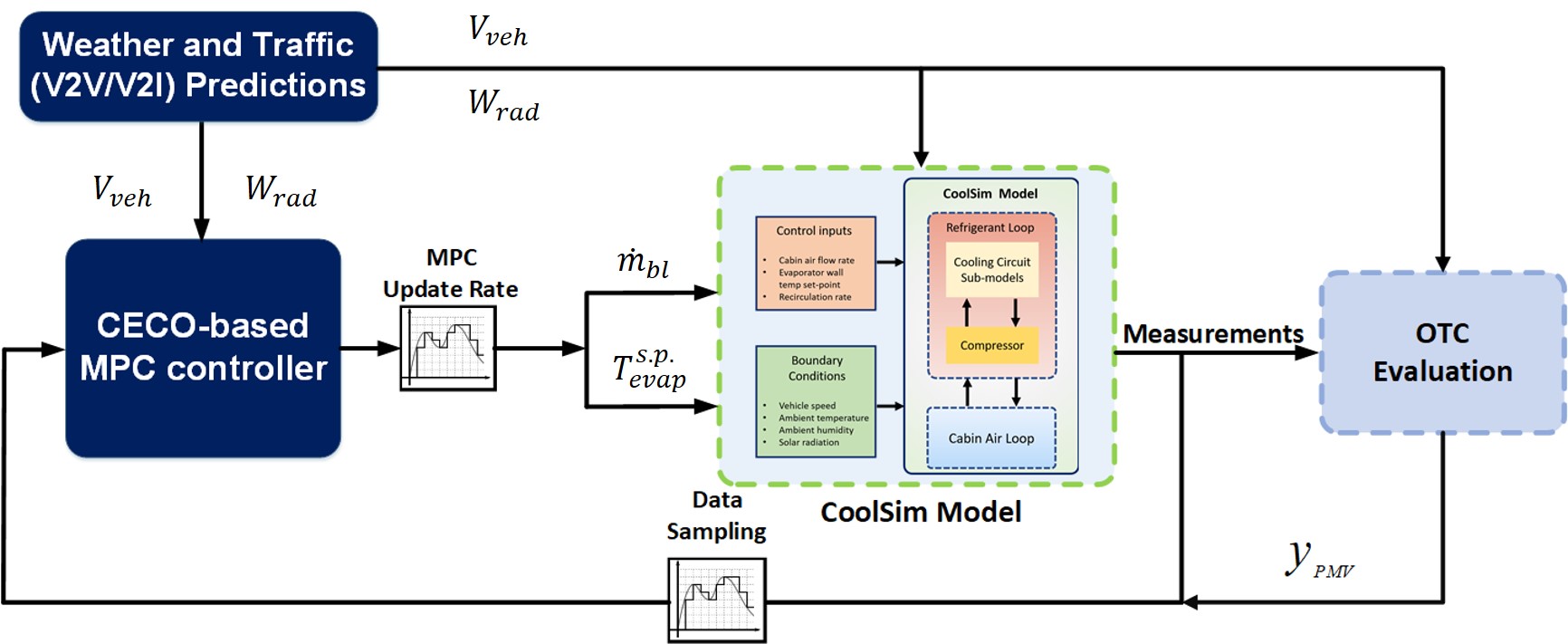}\vspace{-0.2cm}    % The printed column width is 8.4 cm.
		\caption{Schematic of implementing CECO-based MPC controller with CoolSim model in Simulink\textsuperscript{\textregistered}.} 
		\label{fig_ACcontroldiagram}\vspace{-0.7cm}
	\end{center}
\end{figure*}
where $\eta_{AC}\geq 1$ is an efficiency multiplier \cite{Reza19}, which is a function of the vehicle speed (larger value of $\eta_{AC}$ represents higher efficiency in A/C system), $\beta$, $\xi$, and $\epsilon^{UB}$ represent the weighing factor, the regularity term, and the constant upper bound on $\epsilon$, respectively. When compared with the problem formulation in (\ref{eqn:CECO}), an additional optimization variable $\epsilon(i|k)$ is introduced and calculated online to actively tighten $y_{PMV}^{UB}(i|k)$ considering the speed sensitivity of the A/C system efficiency \cite{Hao18}. The basic idea of CECO-IOCH is tightening $y_{PMV}^{UB}(i|k)$ to provide better OTC only when A/C system is operating in high efficiency regions (i.e., high vehicle speed regions). By utilizing this speed sensitivity, better energy efficiency may be achieved while maintaining the same OTC level. This IOCH mechanism was first proposed in \cite{Reza19} for tightening the constraint on $T_{cab}$ based on the same speed sensitivity exploited here. $y_{PMV}^{UB}(i|k)$ and $y_{PMV}^{LB}(i|k)$ used in CECO-IOCH case are the same as the ones applied in the general CECO problem. Note that CECO-IOCH leverages both weather ($W_{rad}$) and traffic ($V_{veh}$) preview information while CECO-E and CECO-C only utilize the weather prediction. The NMPC problems (\ref{eqn:CECO})--(\ref{eqn:CECO-IOCH}) are solved numerically using the MPCTools package \cite{mpctools}. This package exploits CasADi \cite{CAS18} for automatic differentiation and IPOPT algorithm for the numerical optimization.\vspace{-0.5cm}  

\section{Simulation Results}\label{sec:4}
\begin{figure}[h!]
	\begin{center}
		\includegraphics[width=9.0cm]{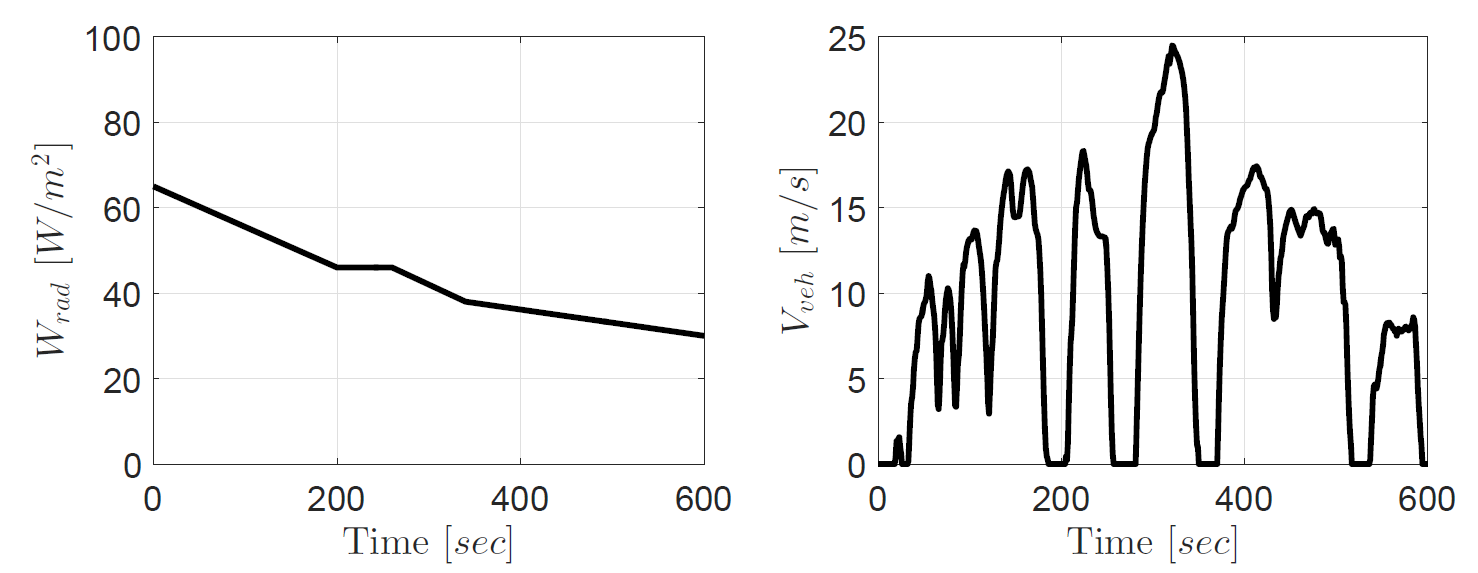} \vspace{-0.7cm}   % The printed column width is 8.4 cm.
	\end{center}\vspace{-0.3cm}  
	\caption{Weather and traffic preview information for the case studies, which are assumed available via CAV technologies.}
	\label{fig_WradVveh_traj}  \vspace{-0.5cm} 
\end{figure}

In order to compare with the CECO-based designs (i.e., CECO-E, CECO-C, and CECO-IOCH), a baseline strategy is defined by applying a PI anti-windup controller to track a constant cabin temperature set-point. This baseline strategy represents a more conventional A/C system control strategy, which considers the average cabin temperature as the only measure of the OTC. The implementation of CECO-based MPC controller in closed-loop with CoolSim model is illustrated in Fig.~\ref{fig_ACcontroldiagram}, which takes the measurements from the CoolSim model, the OTC evaluation, the weather and traffic predictions as inputs and updates the controls of the A/C system. The three design scenarios are implemented by solving variations of the general CECO problem in (\ref{eqn:CECO}). The MPC controller is updated at the sampling time $T_s=5 ~sec$ with prediction horizon $N_p=6$. The weather ($W_{rad}$) and traffic ($V_{veh}$) previews are assumed to be known over the prediction horizon. Their trajectories over SC03 cycle are shown in Fig.~\ref{fig_WradVveh_traj}. \vspace{-0.5cm} 

\begin{figure*}[t]
	\begin{center}
		\includegraphics[scale=0.18]{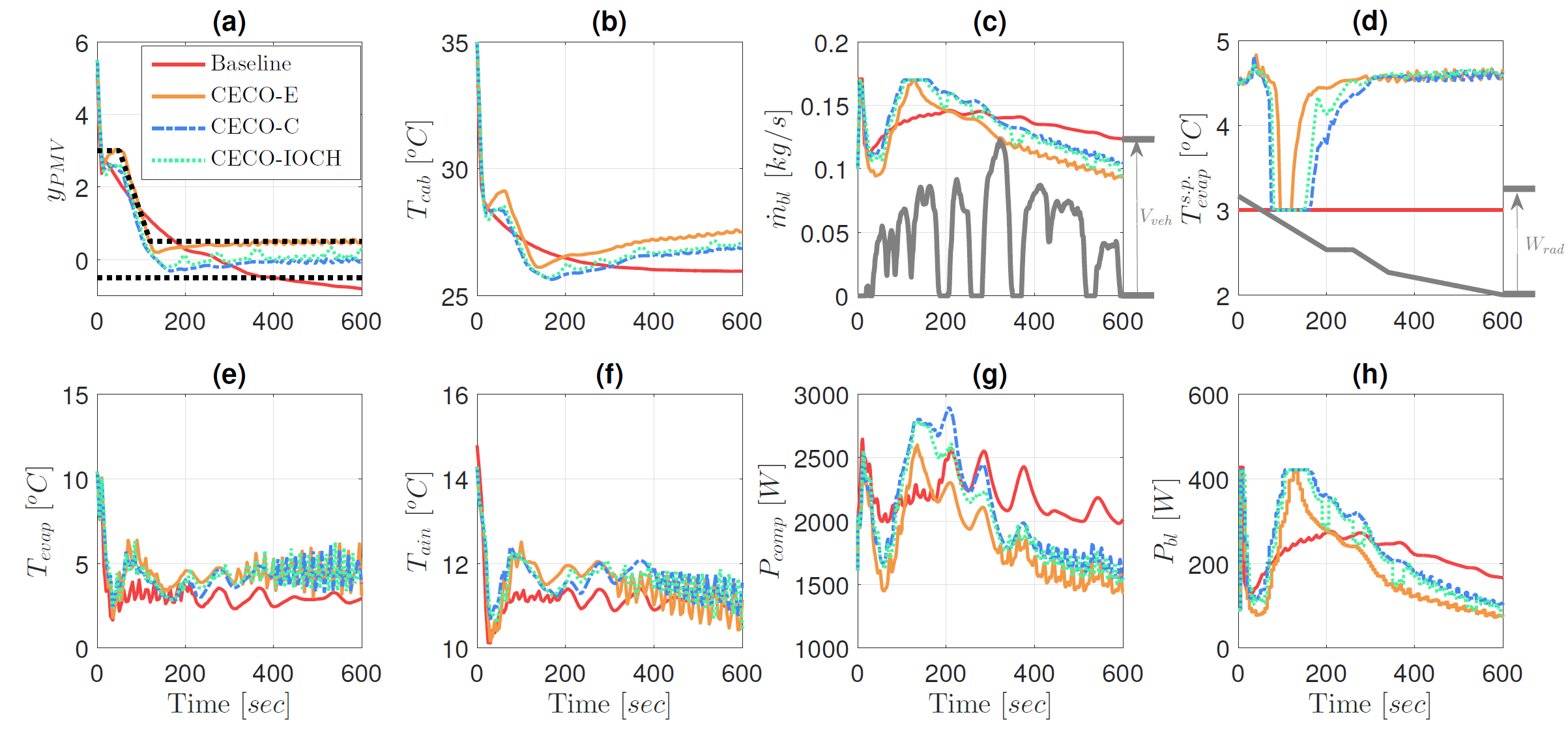}\vspace{-0.2cm}    % The printed column width is 8.4 cm.
		\caption{Simulation results from CoolSim model when comparing different CECO-based designs with the baseline strategy.} 
		\label{fig_CoolSim_outputs_comp}\vspace{-0.7cm}
	\end{center}
\end{figure*}

\begin{figure}[h!]
	\begin{center}
		\includegraphics[width=9.0cm]{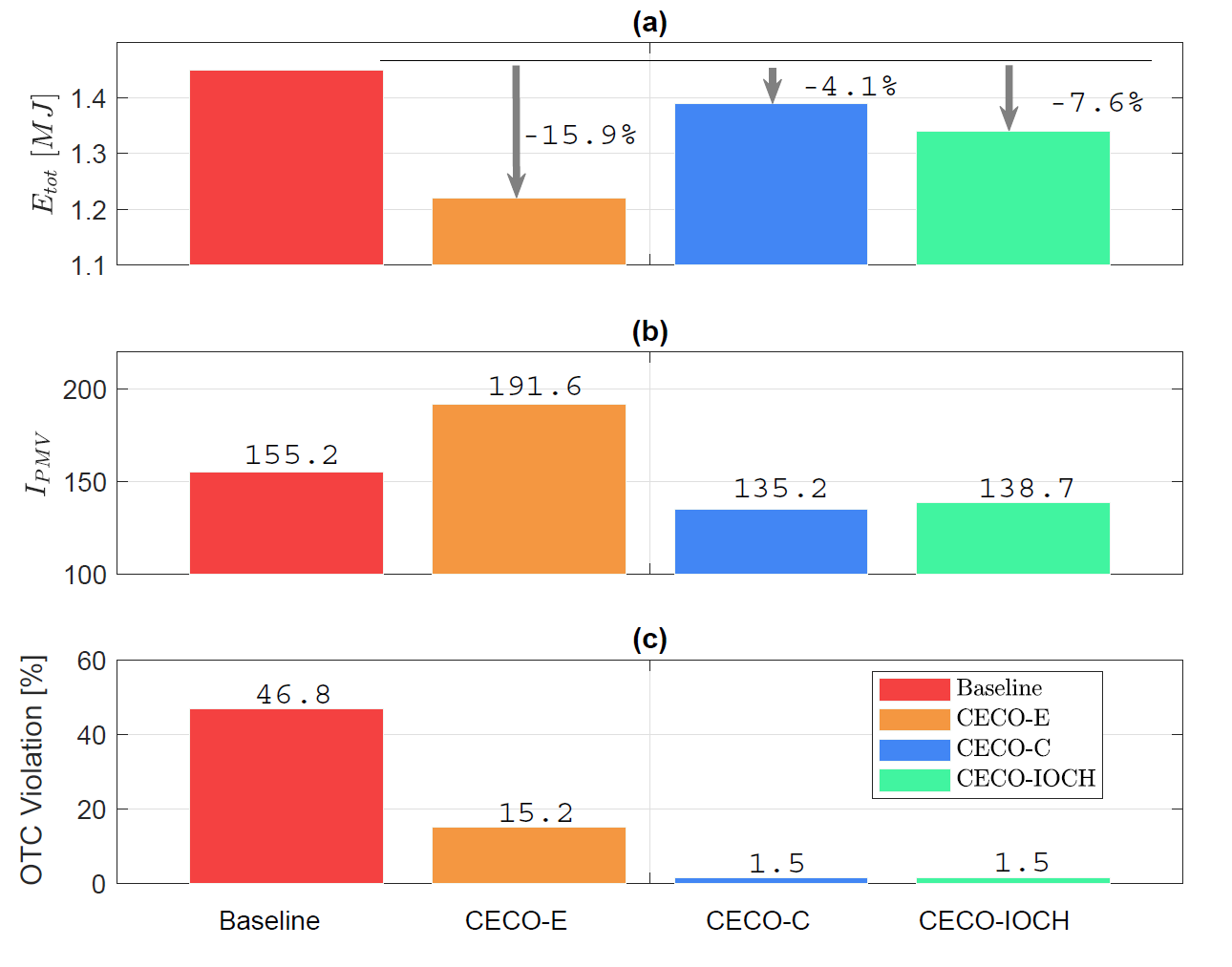} \vspace{-0.5cm}   % The printed column width is 8.4 cm.
	\end{center} \vspace{-0.5cm} 
	\caption{A/C system energy consumption and the OTC comparisons.} \vspace{-0.5cm} 
	\label{fig_Energy_comfort_comp} 
\end{figure}

In Fig.~\ref{fig_CoolSim_outputs_comp}, the time histories of major system outputs based on the closed-loop simulations with the CoolSim model are shown for different control strategies. In Fig.~\ref{fig_CoolSim_outputs_comp}-(a) which shows the $y_{PMV}$ trajectories, the hard constraints on the OTC are plotted in dotted black lines. It is shown that the baseline controller regulates $T_{cab}$ to track the set-point ($26~^oC$), however, according to $y_{PMV}$, this baseline strategy violates the OTC constraints for a significant amount of time, leading to both undercooling and overcooling cases. In comparison, all CECO-based control designs are able to overall maintain the OTC within the comfort zone. The energy and comfort comparisons of these four cases are provided in Fig.~\ref{fig_Energy_comfort_comp}. The total A/C energy consumption ($E_{tot}$) over the simulation time $T$ is calculated by  \vspace{-0.35cm} 
\begin{eqnarray}
E_{tot}=\int_{t=0}^{T} (P_{comp}(t)+P_{bl}(t)) dt.
\end{eqnarray}

\begin{figure}[h!]
	\begin{center}
		\includegraphics[width=8.0cm]{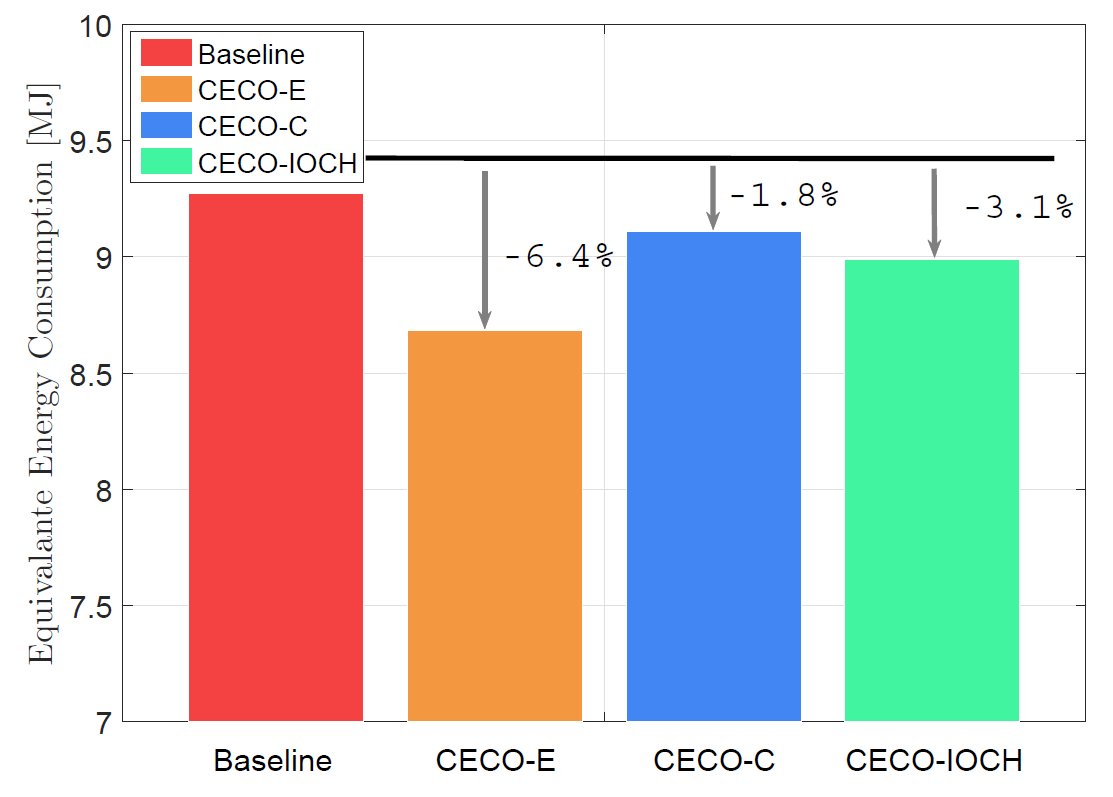} \vspace{-0.3cm}   % The printed column width is 8.4 cm.
	\end{center} \vspace{-0.35cm} 
	\caption{Vehicle-level equivalent energy consumption comparison.} \vspace{-0.5cm} 
	\label{fig_vehicle_fuel} 
\end{figure}

To quantify the OTC level, two metrics are considered: $I_{PMV}$ is defined as \vspace{-0.35cm} 
  \begin{eqnarray}
I_{PMV}=\int_{t=0}^{T} y_{PMV}^2(t) dt,
\end{eqnarray}
and OTC violation is defined as \vspace{-0.35cm} 
  \begin{eqnarray}
\text{OTC violation}=\frac{\tilde{T}}{T}\times 100\%,
\end{eqnarray}
where, \vspace{-0.35cm} 
%\begin{eqnarray}
%\tilde{T}=\Big\{\int_{t=0}^{T} 1 dt~\Big|~ y_{PMV}(t)>y_{PMV}^{UB}(t) \lor y_{PMV}(t)<y_{PMV}^{LB}(t)\Big\}. \nonumber
%\end{eqnarray}
\begin{eqnarray}
\tilde{T}=&\{&\int_{t=0}^{T} x(t) dt~\Big|~ \\\nonumber x(t)&=&\left\{
\begin{array}{ll}
1: \text{if}~ y_{PMV}(t)>y_{PMV}^{UB}(t) ~\text{or}~ y_{PMV}(t)<y_{PMV}^{LB}(t),\\
0: \text{otherwise},
\end{array}
\right. \}. \nonumber
\end{eqnarray}

As shown in Fig.~\ref{fig_Energy_comfort_comp}, comparing CECO-E with the baseline strategy, the energy consumption of the A/C system is reduced by 15.9\%, and lower OTC violation is achieved with higher $I_{PMV}$ value. When CECO-C is applied, 4.1\% energy is saved compared with the baseline strategy, meanwhile, lower values in $I_{pmv}$ and OTC violation indicate that better comfort level has also been achieved. By comparing the CECO-IOCH results with the CECO-C, the benefit of incorporating traffic preview information is demonstrated. While CECO-IOCH design provides similar OTC level, it saves more energy by exploiting the speed sensitivity of the A/C system efficiency. Note that the OTC violations in CECO-C and CECO-IOCH are all from the initial responses at the very beginning of the cycle while in CECO-E, due to model mismatch, operating close to the constraint leads to higher OTC violation which may occur at certain instants over the entire simulation period. CECO-based designs save energy by avoiding the overcooling, which consumes energy and compromises the comfort. In our simulation case studies, as $W_{rad}$ gradually decreases (shown in Fig.~\ref{fig_WradVveh_traj}), the CECO-based designs coordinate with the trend of $W_{rad}$ and decrease the A/C usage accordingly while enforcing the OTC constraints. 

To further validate the impact of the proposed CECO-based strategy on vehicle level energy consumption, we utilize the same powertrain model from \cite{Reza19}, which was developed and partially experimentally validated against a Prius HEV (MY 2017) data. For the validation, the same powertain controller is applied with the A/C power trajectories from different design cases shown in Fig.~\ref{fig_CoolSim_outputs_comp}. As can be seen from Fig.~\ref{fig_vehicle_fuel}, CECO-E, CECO-C, and CECO-IOCH reduce the vehicle energy consumption by $6.4\%$, $1.8\%$, and $3.1\%$, respectively, when compared with the baseline strategy over the SC03 driving cycle.\vspace{-0.5cm}  

\section{Conclusions}\label{sec:5}
A combined energy and comfort optimization (CECO) strategy for the A/C system in connected and automated vehicles (CAVs) has been proposed in this paper with the objective of minimizing the A/C energy consumption and maintaining the occupant thermal comfort (OTC) level. The OTC has been modeled using a modified predictive mean vote (PMV) model which has been modified to account for the special characteristics of the automotive A/C systems. By varying the formulation in the general CECO problem, three CECO-based controllers have been developed and tested on the high-fidelity CoolSim model. These controllers differ in their ways of leveraging the weather and traffic preview information and enforcing the OTC constraints. In the simulations over the SC03 driving cycle, the CECO-based controllers can reduce the A/C energy consumption by up to 7.6\%, which translates into 3.1\% vehicle fuel savings, while providing a better OTC level when compared with a conventional PI and anti-windup based controller which tracks a constant cabin temperature set-point. The trade-off between energy and comfort for different cases has also been highlighted.\vspace{-0.2cm} 
% Here's where you specify the bibliography style file.
% The full file name for the bibliography style file 
% used for an ASME paper is asmems4.bst.

 \vspace{-0.35cm} 
\begin{acknowledgment}
	This work is supported by the United States Department of Energy (DOE), ARPA-E NEXTCAR program (Award No.: DE-AR0000797).
\end{acknowledgment}

 \vspace{-0.35cm} 
\bibliographystyle{asmems4}

%%%%%%%%%%%%%%%%%%%%%%%%%%%%%%%%%%%%%%%%%%%%%%%%%%%%%%%%%%%%%%%%%%%%%%

%%%%%%%%%%%%%%%%%%%%%%%%%%%%%%%%%%%%%%%%%%%%%%%%%%%%%%%%%%%%%%%%%%%%%%
% The bibliography is stored in an external database file
% in the BibTeX format (file_name.bib).  The bibliography is
% created by the following command and it will appear in this
% position in the document. You may, of course, create your
% own bibliography by using thebibliography environment as in
%
% \begin{thebibliography}{12}
% ...
% \bibitem{itemreference} D. E. Knudsen.
% {\em 1966 World Bnus Almanac.}
% {Permafrost Press, Novosibirsk.}
% ...
% \end{thebibliography}

% Here's where you specify the bibliography database file.
% The full file name of the bibliography database for this
% article is asme2e.bib. The name for your database is up
% to you.
%\bibliography{asme2e}

\begin{thebibliography}{99}
\bibitem{Guanetti18}
J. Guanetti, Y. Kim, and F. Borrelli, ``Control of connected and automated
vehicles: state of the art and future challenges,'' {\em Annual Reviews
	in Control}, vol. 45, pp. 18-40, 2018.

\bibitem{Vahidi18}
A. Vahidi and A. Sciarretta, ``Energy saving potentials of connected and automated vehicles," {\em Transportation research Part C}, vol. 95, pp. 822-843, 2018.

\bibitem{Rask2014}
E. Rask, ``Ford focus BEV in-depth (Level 2) testing and
analysis," {\em Presented at Vehicle Systems Analysis Technical	Team (VSATT) meeting}, April, 2014.

\bibitem{Jeffers2015} 
M. Jeffers, L. Chaney, and J. Rugh, ``Climate control load reduction strategies for electric drive vehicles in warm weather," (No. 2015-01-0355) {\em SAE Technical Paper}, 2015.	

\bibitem{Daanen2003}
H.A.M. Daanen, E. Vliert, X. Huang, ``Driving performance in cold, warm, and thermoneutral environments," {\em Applied ergonomics}, vol. 34, no. 6, pp. 597-602, 2003.

\bibitem{HKhayyam2011}
H. Khayyam, A.Z. Kouzani, E.J. Hu, and S. Nahavandi, ``Coordinated energy management of vehicle air conditioning system," {\em Applied thermal engineering}, vol. 31, no. 5, pp. 750-764, 2011.

\bibitem{QZhang2016}
Q. Zhang, S. Stockar, and M. Canova, ``Energy-optimal control of an automotive air conditioning system for ancillary load reduction," {\em IEEE Transactions on Control Systems Technology}, vol. 24, no. 1, pp. 67-80, 2016.

\bibitem{XYan2018}
X. Yan, F. James, and L. Roberto, ``A/C Energy Management and Vehicle Cabin Thermal Comfort Control," {\em IEEE Transactions on Vehicular Technology}, vol. 67, no. 11, pp. 11238-11242, Nov. 2018.

\bibitem{Hao18}
H. Wang, I. Kolmanovsky, M. Amini, and J. Sun, ``Model predictive
climate control of connected and automated vehicles for improved
energy efficiency," in {\em American Control Conference}, June 27-29, 2018,
Milwaukee, WI, USA.

\bibitem{Reza19}
M. Amini, H. Wang, X. Gong, D. Liao-McPherson, I. Kolmanovsky, and J. Sun, ``Cabin and battery thermal management of connected and automated HEVs for improved energy efficiency using hierarchical model predictive control,"
{\em IEEE Transactions on Control Systems Technology}, 2019 (doi: 10.1109/TCST.2019.2923792).

\bibitem{Reza19_2}
M. Amini, X. Gong, Yiheng Feng, H. Wang, I. Kolmanovsky, and J. Sun, ``Sequential Optimization of Speed, Thermal Load, and Power Split in Connected HEVs," in {\em American Control Conference}, July 10-12, 2019, Philadelphia, PA, USA.

\bibitem{Martinho2003}
 N.A.G. Martinho, M.C.G. Silva, and J.A.E. Ramos, ``Evaluation of thermal comfort in a vehicle cabin," {\em Proceedings of the Institution of Mechanical Engineers, Part D: Journal of Automobile Engineering}, vol. 218, no.2, pp. 159-166, 2004.
 
 \bibitem{Hao19}
 H. Wang, Y. Meng, Q. Zhang, M. Amini, I. Kolmanovsky, J. Sun, and M. Jennings, ``MPC-based Precision Cooling Strategy (PCS) for Efficient Thermal Management of Automotive Air Conditioning System," in {\em IEEE Conference on Control Technology and Applications (CCTA)}, August 19-21, 2019, Hong Kong, China.
 
 \bibitem{Hoof08}
 J.V. Hoof, ``Forty years of Fanger's model of thermal comfort: comfort for all," {\em Indoor Air}, vol. 18, no. 3, pp. 182-201, 2008.
 
 \bibitem{ISO7730}
 ISO 7730:2005(en), Ergonomics of the thermal environment - Analytical determination and interpretation of thermal comfort using calculation of the PMV and PPD indices and local thermal comfort criteria.
 
 \bibitem{Croitoru15}
 R. Croitoru, I. Nastase, F. Bode, A. Meslem, and A. Dogeanu, ``Thermal comfort models for indoor spaces and vehicles - Current capabilities and future perspectives,"
 {\em Renewable and Sustainable Energy Reviews}, vol. 44, pp. 304-318, 2015.
 
 \bibitem{Freire08}
 R. Z. Freire, G.H.C. Oliveira, and N. Mendes, ``Predictive controllers for thermal comfort optimization and energy savings," {\em Energy and Buildings}, vol. 40, no.7, pp. 1353-1365, 2008.
 
 \bibitem{Ku15}
 K. L. Ku, J. S. Liaw, M. Y. Tsai and T. S. Liu, ``Automatic Control System for Thermal Comfort Based on Predicted Mean Vote and Energy Saving," {\em IEEE Transactions on Automation Science and Engineering}, vol. 12, no. 1, pp. 378-383, Jan. 2015.
 
 \bibitem{Chen15}
 X. Chen, Q. Wang, and J. Srebric, ``Model predictive control for indoor thermal comfort and energy optimization using occupant feedback," {\em Energy and Buildings}, vol 102, pp. 357-369, 2015.
 
 \bibitem{Farzaneh08}
 Y. Farzaneh, and A. A. Tootoonchi, ``Controlling automobile thermal comfort using optimized fuzzy controller," {\em Applied Thermal Engineering}, vol.28. no.14, pp. 1906-1917, 2008.
 
 \bibitem{Ilya10}
S.P. Szwabowski, P. MacNeille, I. Kolmanovsky, and D. Filev, 
``In-vehicle ambient condition sensing based on wireless internet access," {\em SAE Paper}, 2010-01-0461, SAE World Congress, Detroit, MI, 2010.

 \bibitem{Brusey18}
 J. Brusey, D. Hintea, E. Gaura, and N. Beloe, ``Reinforcement learning-based thermal comfort control for vehicle cabins," {\em Mechatronics}, vol. 50, pp. 413-421, 2018.

\bibitem{SSchaut19}
S. Schaut and O. Sawodny, ``Thermal Management for the Cabin of a Battery Electric Vehicle Considering Passengers' Comfort," {\em IEEE Transactions on Control Systems Technology}, 2019 (doi: 10.1109/TCST.2019.2914888).

 \bibitem{Kiss13}
 T. Kiss, and L. Chaney, ``A New Automotive Air Conditioning System Simulation Tool Developed in MATLAB/Simulink," {\em SAE Int. J. of Passenger Cars-Mechanical Systems}, vol. 6, vo. 2, pp. 826-840, 2013.
 
 \bibitem{QZhang2016_2}
 Q. Zhang, S.E. Li, and K. Deng, ``Automotive air conditioning: optimization, control and diagnosis," Springer, 2016.
 
 \bibitem{mpctools}
 M.J. Risbeck, and J.B. Rawlings, ``MPCTools: nonlinear model predictive control tools for CasADi,'' 2016.
 
 \bibitem{CAS18}
 J. A. Andersson, J. Gillis, G. Horn, J. B. Rawlings, and M. Diehl, ``CasADi -- A software framework for nonlinear optimization and optimal control,"
 {\em Mathematical Programming Computation}, in press, 2018.
\end{thebibliography}

%%%%%%%%%%%%%%%%%%%%%%%%%%%%%%%%%%%%%%%%%%%%%%%%%%%%%%%%%%%%%%%%%%%%%%
%\appendix       %%% starting appendix
%\section*{Appendix A: Head of First Appendix}
%Avoid Appendices if possible.
%
%%%%%%%%%%%%%%%%%%%%%%%%%%%%%%%%%%%%%%%%%%%%%%%%%%%%%%%%%%%%%%%%%%%%%%%
%\section*{Appendix B: Head of Second Appendix}
%\subsection*{Subsection head in appendix}
%The equation counter is not reset in an appendix and the numbers will
%follow one continual sequence from the beginning of the article to the very end as shown in the following example.
%\begin{equation}
%a = b + c.
%\end{equation}

\end{document}